\documentclass{kluwer}    
\usepackage[dvips]{graphicx}
\newdisplay{guess}{Conjecture}

\begin{document}
\begin{article}
\begin{opening}
\title{Evolutionary Synthesis Modelling of Young Star Clusters in Merging Galaxies}
\author{Peter \surname{Anders}, Uta Fritze -- v. Alvensleben}
\runningauthor{Anders, Fritze--v. Alvensleben, \& de Grijs}
\runningtitle{Evolutionary Synthesis Modelling of Young Star Clusters in Merging Galaxies}
\institute{Universit\"ats-Sternwarte G\"ottingen, Geismarlandstr. 11, 37083
G\"ottingen, Germany}
\author{Richard de Grijs}
\institute{Institute of Astronomy, Madingley Road, Cambridge CB3 0HA, UK}
\begin{abstract}
The observational properties of globular cluster systems (GCSs) are
vital tools to investigate the violent star formation histories of their
host galaxies. This violence is thought to have been triggered by
galaxy interactions or mergers. The most basic properties of a GCS are
its luminosity function (number of clusters per luminosity bin) and
color distributions. \\
A large number of observed GCS show bimodal color distributions, which
can be translated into a bimodality in either metallicity and/or age. 
An additional uncertainty comes into play when one considers extinction.\\
These effects can be disentangled either by obtaining spectroscopic data
for the clusters or by imaging observations in at least four passbands. 
This allows us then to discriminate between various formation scenarios of
GCSs, e.g. the merger scenario by Ashman \& Zepf, and the multi-phase
collapse model by Forbes et. al.\\
Young and metal-rich star cluster populations are seen to form in
interacting and merging galaxies. We analyse multiwavelength broad-band
observations of these young cluster systems provided by the ASTROVIRTEL
project. 
\end{abstract}
\end{opening}

\section{Modelling multi-colour star cluster data}

We have further extended the G\"ottingen evolutionary synthesis (ES)
code by including the effects of gaseous emission. The gaseous emission
contributes significantly to the integrated light of stellar populations
younger than $10^8$ years \cite{Andersa}; see Figure 1. In addition,
the effect of {\sl internal} dust extinction has been included. 

The simultaneous determination of a cluster's age, metallicity,
extinction and mass is achieved by comparing an appropriate grid of ES
models with the observed spectral energy distributions (SEDs) in a
least-squares sense. Due to the well-known age-metallicity degeneracy
(and a similar age-extinction degeneracy) at optical wavelengths (and
the dependence on the mass of the cluster), the use of multi-passband
observations is essential to determine these parameters independently. 

\begin{figure}[ht]
\includegraphics[angle=-90,width=0.5\columnwidth]{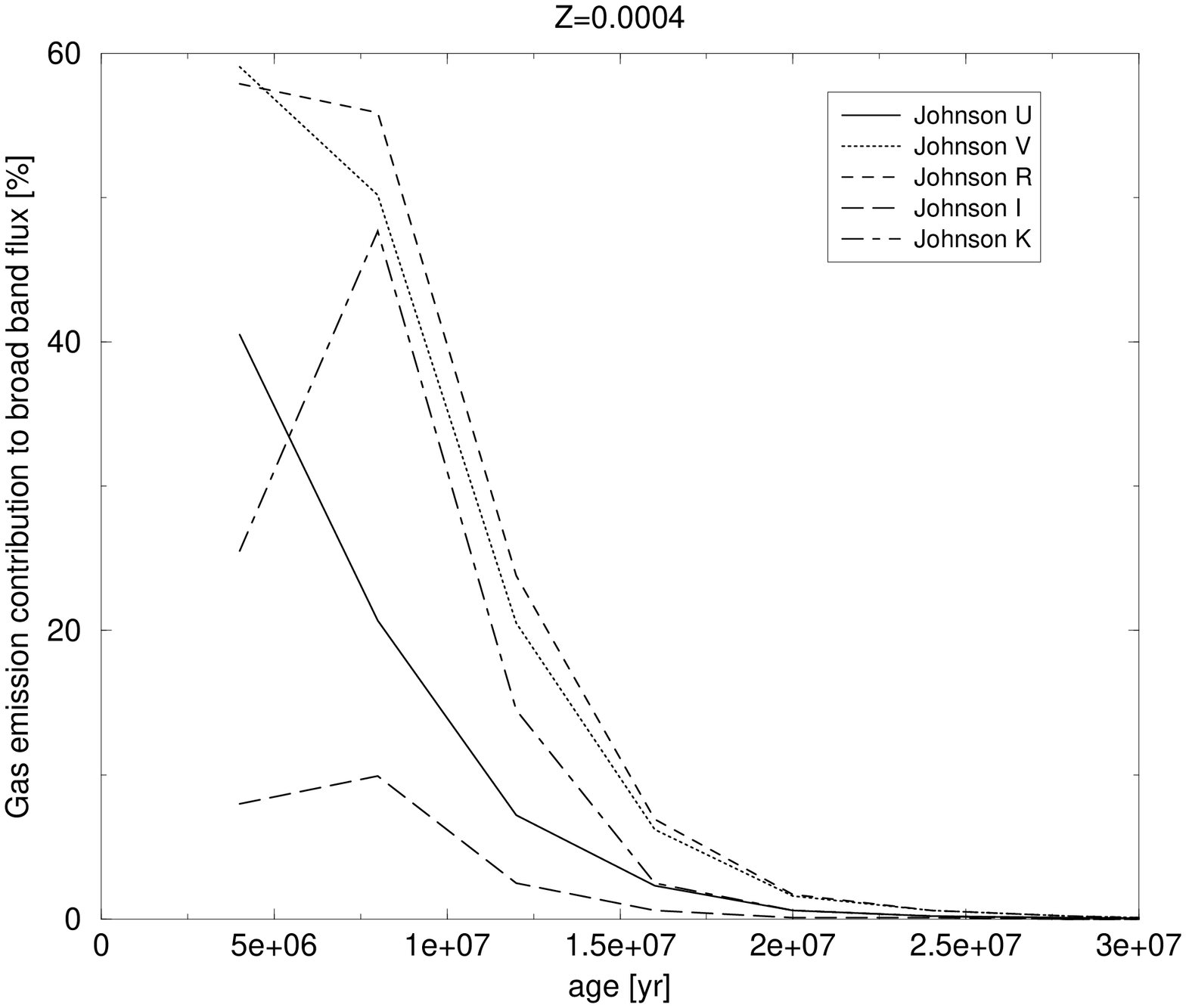}
\includegraphics[angle=-90,width=0.5\columnwidth]{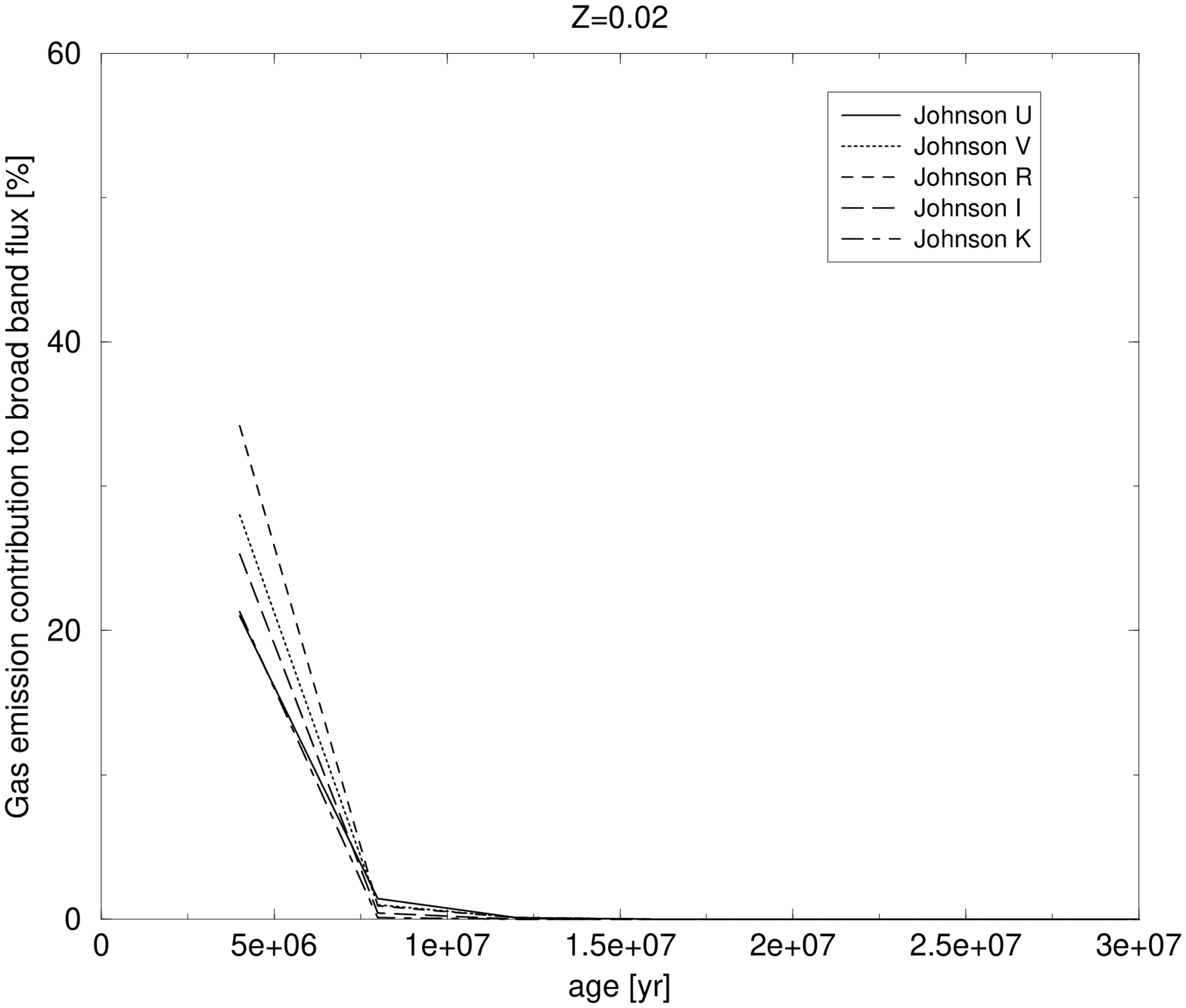}
\caption{Time evolution of the gaseous emission contribution to
broad-band fluxes in Johnson passbands $U, B, V, I$, and $K$ at low
metallicity ${\rm Z = 0.0004}$ (left panel) and solar metallicity (right
panel).}
\end{figure}

\subsection{Results for Artificial Clusters}

We constructed five artificial clusters with known properties, i.e.,
solar metallicity, extinction $E(B-V)=0.1$, ages of 8, 60, and 200 Myr,
and 1 and 10 Gyr. The SEDs for these artificial clusters were taken
from our model grid, and a typical observational error of 0.1 mag was
assumed. Subsequently, we redetermined the best-fit cluster parameters
using our ES code, and compared them to the original values. We
constructed a number of additional clusters characterized by color
offsets (too red or too blue by 0.1 mag) to assess the accuracy of our
model fits. From this study we conclude that

\begin{enumerate}

\item The choice of passband combination required for a proper
determination of the basic cluster parameters (age, metallicity,
extinction) can be optimized towards the expected mean parameter values. 

\item In order to determine reliable ages, the use of ultraviolet (UV)
passbands is vital for all ages. Including a near-infrared (NIR)
passband improves the accuracy further. 

\item For metallicity determinations, NIR observations are most
important, while the UV is essential for young systems (ages $\le$ 100 Myr). 

\item The extinction is best determined using UV + optical colors.

\end{enumerate}

In general, the accuracy and the best passband combination are a strong
function of the age of the cluster's stellar population \cite{Andersb}. 
The availability of observations spanning the entire wavelength range,
from UV to NIR, allows to constrain all parameters (age, metallicity and
internal extinction) most efficiently. 

\section{Application and first results: NGC 1569}

The irregular dwarf galaxy NGC 1569 is commonly classified as a
(post) starburst galaxy with H{\sc ii} region metallicity $\sim$
(0.2 -- 0.5) $Z_\odot$. High-resolution multipassband imaging data were
provided by the ESO / ST-ECF ASTROVIRTEL project ``The Evolution and
Environmental Dependence of Star Cluster Luminosity Functions'' (PI R. 
de Grijs). In addition to the two well-studied super star clusters in
its main disk, they reveal a number of fainter objects resembling
compact star clusters (see also \cite{Hunter}).

Our fitting algorithm provides evidence for continuous star formation
starting at least 7 Gyr ago. The recent burst, however, started around
80 Myr ago, with a possible peak of cluster formation around 35 Myr ago
and a major peak in the cluster formation rate in the youngest age bin
$\le$ 8 Myr (see Figure 2). These results are consistent with previous
results \cite{Greggio,Aloisi} regarding the bulk of the starburst
activity in general and the formation of the two super star clusters in
particular. As seen in previous studies, the most recent, enhanced
epoch of star formation coincides well with the formation of
morphologically peculiar features, like H$\alpha$ filaments and
``superbubbles'' \cite{Waller}. 

\begin{figure}[!ht]
\includegraphics[width=0.5\columnwidth]{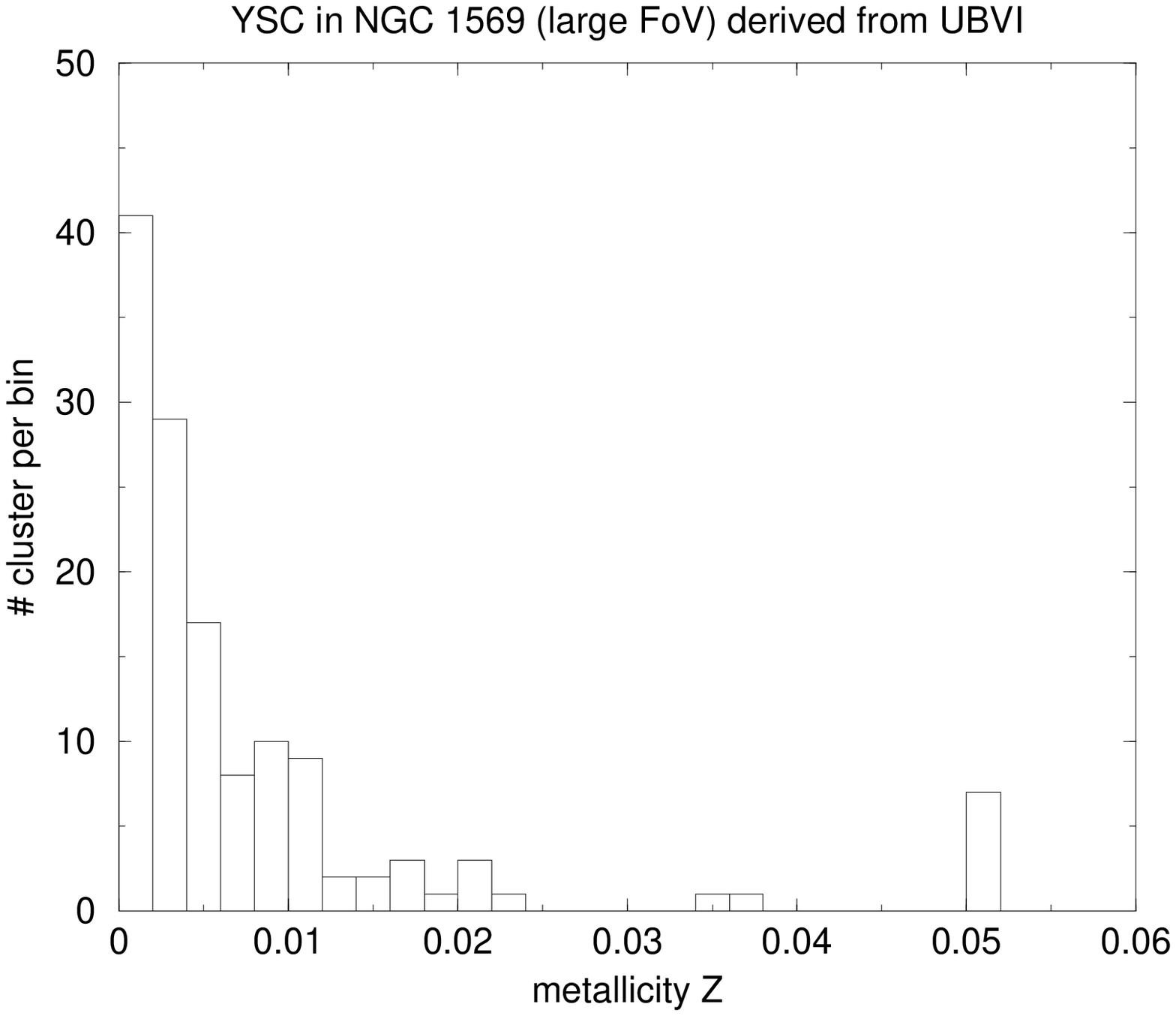}
\includegraphics[width=0.5\columnwidth]{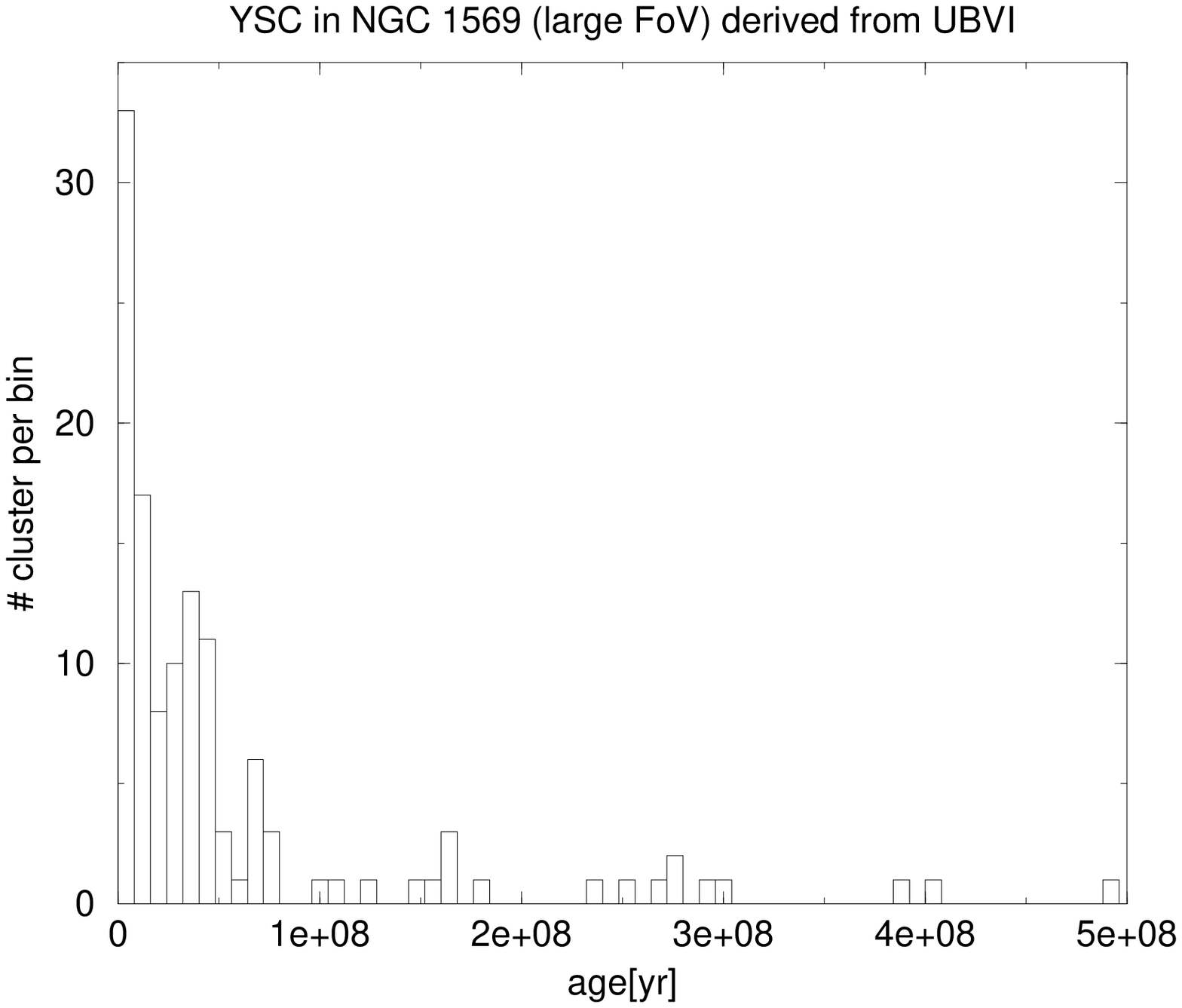}
\caption{Left panel: Metallicity distribution of young star clusters in NGC 1569,
derived from $U, B, V$ and $I$; right panel:
corresponding age distribution (during the burst; nine older clusters,
up to 7 Gyr old, are also present).}
\end{figure}

An accompanying H{\sc i} cloud is located at 5 kpc from NGC 1569,
connected to it by an H{\sc i} bridge, suggesting a tidal interaction as
a possible cause for the starburst \cite{StilIsrael}. 

We determine masses for the individual star clusters on the basis of our
model mass-to-light ratios at their respective ages and metallicities. 
We find masses up to $10^6 M_\odot$. The metallicity distribution is
broader than found in previous surveys, which mainly focussed on the ISM
\cite{Kobulnicky,Devost}. This may imply different chemical enrichment
processes in the star clusters vs. the ISM, or selection effects caused
by the different spatial coverage. The internal dust extinction is
found to be low ($E(B-V) < 0.25$, with the majority of the clusters
being affected by $E(B-V) < 0.1$) \cite{Andersc}. 

\acknowledgements
Very efficient support given by ASTROVIRTEL, a Project funded by the
European Commission under FP5 Contract No. HPRI-CT-1999-00081, is
gratefully acknowledged. P. Anders is partially supported by DFG grant
DFG Fr 916/11.

\end{article}
\end{document}